\documentclass[useAMS,usenatbib]{mn2e}
\usepackage{graphicx,epsfig,epsf,amssymb}

\title[Kinematics and chemistry of the hot core in G20.08-0.14N]{Kinematics and chemistry of the hot core in G20.08-0.14N}
\author[Jin-Long Xu  and Jun-Jie Wang]{Jin-Long Xu$^{1,2}$\thanks{E-mail:
xujl@bao.ac.cn} and Jun-Jie Wang$^{1,2}$\\
$^{1}$National Astronomical Observatories, Chinese
Academy of Sciences, Beijing 100012, China\\
$^{2}$NAOC-TU Joint Center for Astrophysics, Lhasa 850000, China\\}
\begin{document}


\pagerange{\pageref{firstpage}--\pageref{lastpage}} \pubyear{2002}

\maketitle

\label{firstpage}

\begin{abstract}
We present Submillimeter Array observations of the massive
star-forming region G20.08-0.14N at 335 and 345 GHz. With the SMA
data, 41 molecular transitions were detected related to 11 molecular
species and their isotopologues, including $\rm SO_{2}$, $\rm
SO$, $\rm C^{34}S$, $\rm NS$, $\rm C^{17}O$, $\rm SiO$, $\rm
CH_{3}OH$, $\rm HC_{3}N$, $\rm H^{13}CO^{+}$, $\rm HCOOCH_{3}$ and
$\rm NH_{2}CHO$.  In G20.08-0.14N, 10 transition lines of the
detected 41 transition lines  belong to SO$_{2}$, which dominates
the appearance of the submillimeter-wave spectrum. To obtain the
spatial kinematic distribution of molecules in G20.08-0.14N, we
chose the strongest and unblended lines for the channel maps. The
channel maps of $\rm C^{34}S$ and $\rm SiO$, together with their
position-velocity diagrams, present that there are two accretion
flows in G20.08-0.14N. Additionally, SiO emission shows a collimated
outflow at the NE-SW direction. The direction of the outflow is for
the first time revealed. The rotational temperature and the column
density of $\rm CH_{3}OH$ are 105 K and 3.1 $\times$ $10^{17}$ $\rm
cm^{-2}$, respectively.  Our results confirm that a hot core is
associated with G20.08-0.14N. The hot core is heated by a protostar
radiation at it center, not by the external excitation from shocks.
The images of the spatial distribution of different species have
shown that the different molecules are located at the different
positions of the hot core. Through comparing the spatial
distributions and abundances of the molecules, we discuss possible
chemical processes for producing the complex sulfur-bearing,
nitrogen-bearing and oxygen-bearing molecules in G20.08-0.14N.
\end{abstract}


\section{Introduction}

Massive stars are formed in dense molecular clouds. Massive star
formation also has a significant effect on the chemistry of the
surrounding molecular clouds \citep{van dishoeck98}. Hot cores are
the formation sites of massive stars. The hot cores are defined as
compact ( $\leq$ 0.1 pc, n $\geq$ $10^{7}$ $\rm cm^{-3}$),
relatively high temperature ($T\rm_{k}$ $\geq$ 100 K) cloud cores
\citep{kurtz00}, whose phase is thought to last about 10$^{5}$ yr
\citep{van dishoeck98}. When the hot core is formed, the central
massive protostar can produce the ionizing radiation and the
associated outflow can produce shocks, hence the hot cores represent
the most chemically rich phase of the massive star formation often
associated with UC HII regions
\citep{Cesaroni92,Hatchell98,Garay99,Churchwell02}. The high
abundances of the organic molecules in hot cores are consequently
attributed to grain-surface chemistry and mantle evaporation
processes \citep{van dishoeck98,Liu01}. Because of the compact and
dense nature of hot cores, single-dish observations with large beam
sizes are not sufficient to explore the dense cores and detailed
kinematics. The interferometer observations at submillimeter wave
can filter out the extended diffuse components, then the detailed
dynamical processes and chemical conditions of hot cores can be
revealed.  A number of line observations at
submillimeter/millimeter wavelengths were previously used to explore
the molecular composition of hot cores
\citep{Beuther05,goddi09,qin10}, but more sources need to be added
to the inventory of studied hot cores before their chemical
evolution can be understood.

G20.08-0.14N is a massive star-forming region. It is approximately
at a distance of 12.3 kpc \citep{Fish03,Anderson09}, corresponding
to a bolometric luminosity of about
6.6$\times$10$^{5}L_{\odot}$\citep{Gal09}. Previous radio continuum
observations at centimeter wavelengths suggest that G20.08-0.14N has
three UC and HC HII regions \citep{Wood89}. In addition, single-dish
observations of molecular lines show the signatures of infall,
accretion, and outflow \citep{Klaassen07}. Moreover, $\rm
H_{2}O$ \citep{Hofner96}, OH \citep{Ho83}, $\rm CH_{3}OH$
\citep{Walsh98} and NH$_{3}$ \citep{Gal09} masers in G20.08-0.14N
were revealed in some observations. These signatures indicate active
massive star formation in this region. The observed CH$_{3}$CN
transitions indicate that a hot molecular core is associated with
G20.08-0.14N \citep{Gal09}. Thus, G20.08-0.14N provides us with an
opportunity to study the physical and chemical conditions of massive
star-forming processes.

We have carried out multiline observations toward the massive
star-forming region G20.08-0.14N with the Submillimeter Array.
Various molecular lines are used to investigate the physical and
chemical processes of G20.08-0.14N.  In Section 2, we summarize the
observations and data reduction. In Section 3, we give the general
results. In Section 4, we present the data analysis for deriving
rotation temperatures, column densities, and abundances of various
species relative to H$_{2}$ and implications for chemistry. In
Section 5, we summarize our main conclusions.

\section{OBSERVATIONS AND DATA REDUCTION}

Observations toward G20.08-0.14N were carried out with the SMA on
2009 May 20, at 335 (lower sideband) and 345 GHz (upper sideband).
The data are from SMA archive\footnote{
http://www.cfa.harvard.edu/rtdc/data/search.html}. The two sidebands
of the SMA covered frequency ranges of 335.6--337.6 GHz and
345.6--347.6 GHz, respectively. The total observing time is 9.38
hours. The phase track center was  R.A.(J2000.0)= $\rm 18
^{h}28^{m}10^{s}.30$ and Dec.(J2000.0) = $\rm -11
^{\circ}28^{\prime}47^{\prime\prime}.8$. The typical system
temperature was 186 K. The spectral resolution is 0.812 MHz,
corresponding to a velocity resolution of 0.7 km $\rm s^{-1}$.  The
bright quasar 3C273 was used for bandpass calibration, while
absolute flux density scales were determined from observations of
Callisto (15 Jy). QSO 1733-130 and QSO 1751+096 were observed for
antenna gain corrections. The calibration and imaging were performed
in Miriad. The continuum image was constructed from the line-free
channels. The spectral cubes were constructed using the
continuum-subtracted spectral channels. Self-calibration was
performed on the continuum data. The gain solutions from the
continuum were applied to the line data. The synthesized beam size
of the continuum  was approximately $2^{\prime\prime}.02$ $\times$
$1^{\prime\prime}.15$ with a P.A. =$72^{\circ}.0$.

\section{RESULTS}
\subsection{Continuum Emission at 0.9 mm}
Figure 1 shows  the 0.9 mm continuum map of G20.08-0.14N obtained by
the SMA observations. Galv\'{a}n-Madrid et al. (2009) resolved the
G20.08-0.14N system into three components with the VLA at 1.3 cm.
Each component represents an HII region marked in our Figure 1
\citep{Wood89}. Particularly, the HII  region A is the brightest and
closest to the peaks at 1.3 mm and 0.9 mm. By using a
two-dimensional Gaussian fit for the continuum emission, we obtained
that the total flux density is 2.72$\pm$0.03 Jy, the deconvolved
source size is 1.83$^{\prime\prime}$ $\times$ 1.22$^{\prime\prime}$
(P.A.=72$^{\circ}$), the peak position is R.A.(J2000)=18$\rm
^{h}$28$\rm ^{m}$10.$\rm ^{s}$307
($\Delta$R.A.=$\pm$0.01$^{\prime\prime}$),
decl.(J2000)=--11$^{\circ}$28$^{\prime}$47.$^{\prime\prime}$846
($\Delta$decl.=$\pm$0.01$^{\prime\prime}$) with an intensity of 1.33
$\pm$ 0.02 Jy beam$^{-1}$. The peak position of the 0.9 mm continuum
emission is coincident with HII region A within the uncertainty. An
$\rm H_{2}O$ maser has been detected  in G20.08-0.14N
\citep{Hofner96}. In Figure 1, the $\rm H_{2}O$ maser is shown with
the black filled triangle. The maser is associated with 0.9 mm
continuum emission, but offset from its the peak position.

\begin{figure}
\vspace{-4mm}
\includegraphics[angle=270, scale=.42]{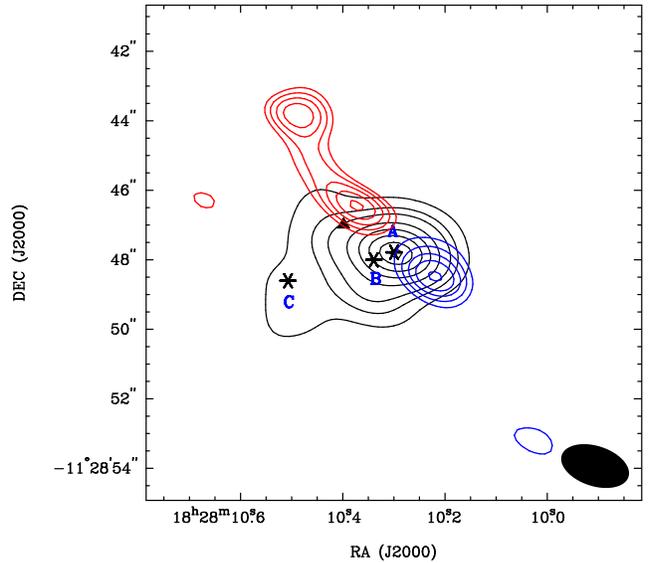}
\vspace{-8mm} \caption{Continuum map toward G20.08-0.14N at 0.9 mm.
The contours are at -4, 4, 8, 12, 20, 28, 36, 44 and 49 $\sigma$.
The rms noise level is 0.03 $\rm Jy\ beam^{-1}$($1\sigma$). The
synthesized beam ($2.0^{\prime \prime}\times1.2^{\prime \prime}$)
with P.A.= 72$^{\circ}$ is shown in the lower right corner.
$``\ast"$ indicates the position of three HII region (Wood \&
Churchwell 1989; Galv\'{a}n-Madrid et al. 2009). $\rm H_{2}O$ maser
is shown with blue filled triangle.  The velocity component of
the blue contours is from 32.5 km s$^{-1}$ to 36.0 km s$^{-1}$,
while the velocity component of the red contours is from 47.0 km
s$^{-1}$ to 51.7 km s$^{-1}$, whose levels are the 20\%, 40\%, 60\%,
80\%, and 100\% of the peak values. }
\end{figure}

\begin{figure*}
\vspace{7mm}
\includegraphics[angle=270,scale=.65]{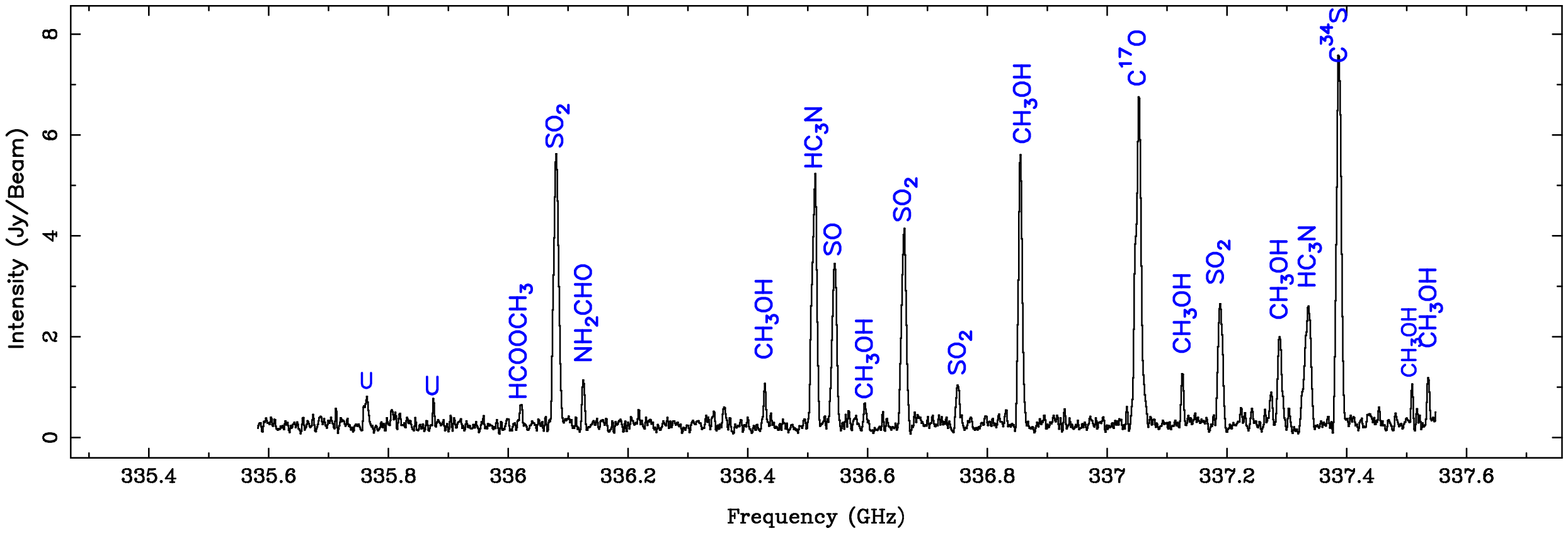}
\includegraphics[angle=270,scale=.65]{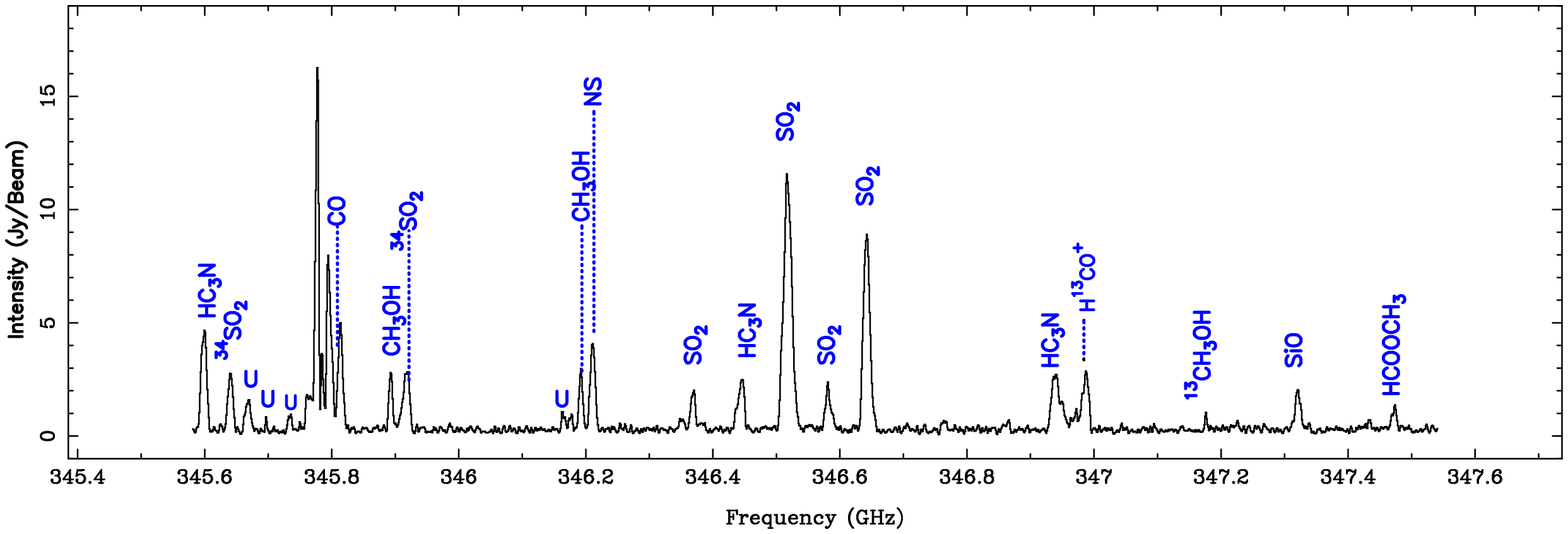}
\vspace{-12mm}\caption{SMA spectra extracted from the line
data-cubes in the image domain toward the G20.08-0.14N. The spectra
are averaged over one beam centered on the peak of the continuum. U
marks the unidentified lines.}
\end{figure*}

\subsection{Molecular Line Emission}

Molecular lines were identified following spectral-line catalogs:
(1) Cologne Database for Molecular Spectroscopy \citep{mul05}, (2)
Molecular Spectroscopy database of Jet Propulsion Laboratory
\citep{Pickett98}, and (3) SPLATALOGUE line catalogs\footnote{
http://www.splatalogue.net} \citep{Remijan07}.

The obtained SMA 4 GHz spectrum is shown in Figure 2 and a full list
of identified lines presented in Table 1. We detected a total of 41
transitions, which include 35 transitions from 11 species and their
isotopologues, and  6 unidentified transitions marked with U. The
identified species contain simple linear molecules as well as
complex oxygen-bearing, nitrogen-bearing, and sulfur-bearing
molecules. Many of the strongest lines in the spectrum are from
diatomic molecules of the most abundant elements. The $^{12}$CO
J=3-2 line at 345.938 GHz is the strongest line in our passband, but
its spectral profile displays absorption features as well. Because
the CO absorption features are at the same velocities as the HI
absorption features of \cite{Fish03}, \cite{Gal09} explained that
these features may be caused by foreground gas that is not
associated with G20.08-0.14N but rather to intervening Galactic
spiral arms. In addition, sulfur-bearing molecules such as SO$_{2}$
and C$^{34}$S display also strong emission lines. The line
parameters are presented in Table 1. The first column is the name of
molecular species. The second and third columns list the transition
and rest frequency of the molecules. The fourth column lists the
upper level energy of each transition, and the fifth column is the
product of the line strength and the square of the relevant dipole
moment. The sixth, seventh, and eighth columns list  the central
line velocity, the peak intensity, and full width at half-maximum
(FWHM) derived from Gaussian fitting to the line profiles,
respectively. The 1 $\sigma$ integrated-intensity noise level of
each transition is given in the ninth column.

\begin{table*}
\begin{center}
\tabcolsep 0.4mm\caption{Observed Parameters of Each Line}
\begin{tabular}{lcccccccccc}
\hline\hline
Molecule   & Transition      & Rest Frequency  & $E_{\rm u}$  & $S\mu^{2}$ & $V_{\rm_{LSR}}$  & $I_{\rm p}$    & $\Delta V$ & Integrated-intensity rms &Notes\\
         &           &(MHz)     &{\rm (K)}   & (debye$^{2}$)&${\rm (km\ s^{-1})}$ &($\rm Jy\ beam^{-1})$& (${\rm km\ s^{-1})}$ & (Jy $\rm beam ^{-1}$ km $\rm s^{-1}$)& \\
  \hline\noalign{\smallskip}
$\rm SO_{2}$    & $23_{_{3,21}}$-$23_{_{2,22}}$ &336089.23  & 276.0 &28.41& 41.8(0.1)   & 4.5(0.1) & 8.8(0.1) &   &    \\   
                & $16_{_{7,9}}$-$17_{_{6,12}} $ &336669.58  & 245.1 &4.34& 41.7(0.1)   & 3.5(0.1) & 8.9(0.2) &   &    \\   
                & $16_{_{4,12}}$-$16_{_{3,13}}$ &346523.86  & 164.5  &23.10& 38.5(0.1)   & 7.8(0.1) & 13.5(0.2) &   &   1 \\   
                & $19_{_{1,19}}$-$18_{_{0,18}}$ &346652.16  & 168.1  &41.98& 41.6(0.1)   & 7.0(0.1) & 10.1(0.1) & 8.0  &    \\   
                & $20_{_{1,19}}$-$19_{_{2,18}}$$v_{_{2}}$=1 &336760.70  & 962.1 &21.72& 42.2(0.2)   & 1.2(0.1) & 7.2(0.4) &   &  2  \\   
                & $12_{_{2,11}}$-$11_{_{1,10}}$$v_{_{2}}$=1 &337191.5   & -- &-- & 35.9(0.1)   & 2.2(0.1) & 7.9(0.2) &  &   3 \\   
                & $19_{_{1,19}}$-$18_{_{1,18}}$$v_{_{2}}$=1 &346379.19  & 930.6  &43.22& 42.0(0.2)   & 1.6(0.1) & 9.4(0.3) &   &    \\   
                & $18_{_{4,14}}$-$18_{_{3,15}}$$v_{_{2}}$=1 &346591.78  & 960.8  &27.65& 41.9(0.1)   & 1.9(0.1) & 9.2(0.3) &   &   4 \\   

$\rm CH_{3}OH$  &$14_{_{7,8}}A^{+}$-$15_{_{6,9}}A^{+}$  &336438.22 & 488.2 &2.38& 42.3(0.3)   & 0.6(0.1) & 4.9(0.5) &  &   \\  
                & $12_{_{1,11}}A^{-}$-$12_{_{0,12}}A^{+}$  &336865.15 & 197.1&22.88& 42.6(0.1)   & 4.8(0.1) & 6.5(0.1)& 3.2  &   \\  
                &$3_{_{3,0}}A^{+}$-$4_{_{2,2}}A^{+}$       &337135.86 & 61.6&0.25& 42.2(0.2)    & 1.2(0.1) & 6.7(0.4) &   & \\   
                & $16_{_{1,15}}A^{-}$-$15_{_{2,14}}A^{-}$ &345903.97  & 332.5  &7.13& 42.1(0.1)   & 2.4(0.1) & 6.0(0.2) &   &    \\   
                & $5_{_{4,2}}A^{-}$-$6_{_{3,3}}A^{-}$      &346202.77      & 115.2  &0.50& 41.6(0.1)   & 2.3(0.1) & 6.5(0.2) &   &  \\ 
                &$7_{_{1,7}}A^{+}$-$6_{_{1,6}}A^{+}$$v_{_{t}}$=1       &337297.44 & 390.0&5.55& 41.8(0.1)    & 1.7(0.1) & 6.4(0.3) &   & \\   
                & $7_{_{3,5}}A$-$6_{_{3,4}}A$$v_{_{t}}$=1 & 337519.1 & 323.9 &4.6& 36.2(0.8)   & 1.0(0.2)   & 6.4(1.8) &  &    \\
                &$7_{_{5,3}}A^{+}$-$6_{_{5,2}}A^{+}$$v_{_{t}}$=1       &337546.05 & 485.4&2.74& 42.1(0.2)    & 1.0(0.1) & 4.2(0.3) &   & 5 \\   
$\rm HC_{3}N$   & $37$-$36$                      &336520.08  &306.9 &55.25& 41.6(0.1)    & 4.2(0.1)   & 8.4(0.1) & 2.0  &  \\  
                & $38$-$37$                     &345609.02  &323.5 &529.13& 41.7(0.1)    & 3.8(0.1)   & 8.5(0.2) &   &  \\  
                & $37$-$36e$                      &337335.35  &1025.3&513.34& 35.4(0.2)    & 2.4(0.1)   & 9.9(0.4) &   &  6 \\ 
                & $38$-$37e$                      &346455.73  &645.1 &526.17& 42.7(0.2)    & 2.2(0.1)   & 11.3(0.4) &   &  \\  
                & $38$-$37f$                      &346949.12  &645.6 &526.19& 40.3(0.2)    & 2.4(0.1)   & 15.6(0.4) &   & 7 \\  

$\rm HCOOCH_{3}$& $27_{_{9,19}}A$-$26_{_{9,18}}A$&336028.15 & 277.4 &64.03& 39.7(0.5)   & 0.4(0.1)   & 7.4(1.0) &  &    \\  
                & $27_{_{5,22}}E$-$26_{_{5,21}}E$&347478.24 & 247.3 &69.17& 37.3(0.2)   & 1.2(0.1)   & 9.1(0.5) & 1.1  &    \\

$\rm ^{34}SO_{2}$ & $5_{_{4,2}}$-$5_{_{3,3}}$    &345651.30 & 51.8 &4.49& 42.5(0.1)   & 2.5(0.1) & 7.7(0.2) &  &    \\  
                & $17_{_{4,14}}$-$17_{_{3,15}}$ &345929.29 & 178.8 &24.49& 43.1(0.1)   & 2.6(0.1) & 11.0(0.3) & 2.0 &    \\  

$\rm NH_{2}CHO$ & $16_{_{2,15}}$-$15_{_{2,14}}$ &336136.88 &149.7 &617.20& 43.5(0.2)   & 1.0(0.1) & 4.8(0.4) & 0.5 &     \\  

$\rm SO$        & $10_{_{11}}$-$10_{_{10}}$     &336553.75 &142.8 &0.29& 41.8(0.1)   & 2.7(0.1) & 8.2(0.2) & 2.1 &    \\  

$\rm C^{17}O$   & $3$-$2$                       &337061.10 & 32.4 &0.04& 41.7(0.1)   & 2.7(0.1) & 12.1(0.2) & 5.2 &    \\  

$\rm C^{34}S$   & $7$-$6$                       &337396.46 & 50.2 &25.57& 41.8(0.1)   & 5.1(0.1) & 7.2(0.1) & 5.2 &    \\  

$\rm CO$        & $3$-$2$                        &345795.99 & 33.2 &0.04& --  & -- & -- & -- &  8  \\  

$\rm NS$        & $15/2,17/2-13/2,15/2f $      & 346221.16 & 70.9 &24.03& 41.7(0.1)   & 3.6(0.1) & 7.2(0.2) & 2.6 &   \\  

$\rm H^{13}CO^{+}$        & $4$-$3$             & 346998.34 & 41.6 &60.85& 42.0(0.2)   & 1.6(0.1) & 7.7(0.3) & 2.8 &  9  \\  

$\rm ^{13}CH_{3}OH$  &  $14_{_{1,13}}A^{-}$-$14_{_{0,14}}A^{-}$  & 347188.28  & 254.3 &25.94& 42.7(0.3)   & 0.7(0.1) & 4.5(0.6) & 0.4 &    \\  
$\rm SiO$  &  $8$-$7$                           & 347330.63 & 75.0 &76.79& 41.0(0.2)   & 1.4(0.1) & 8.7(0.2) & 2.1 &    \\  

\hline
\end{tabular}
\end{center}
Notes.(1) Blend with SO at 346528.5. (2) Questionable identification
(3) Blend with $^{33}$SO at 337195.0. (4) Blend with $^{33}$SO at
336590.5. (5) Blend with $^{34}$SO at 351257.2 (6) Blend with
CH$_{3}$CH$_{2}$CN at 337347.6. (7) Blend with CH$_{3}$CH$_{2}$CN at
346947.3. (8) Very broad feature with three peaks.  (9) Blend with
CH$_{3}$CH$_{2}$CN at 346983.8. \vspace{10mm}
\end{table*}

\begin{figure*}
\vspace{0mm}
\includegraphics[angle=270,scale=.95]{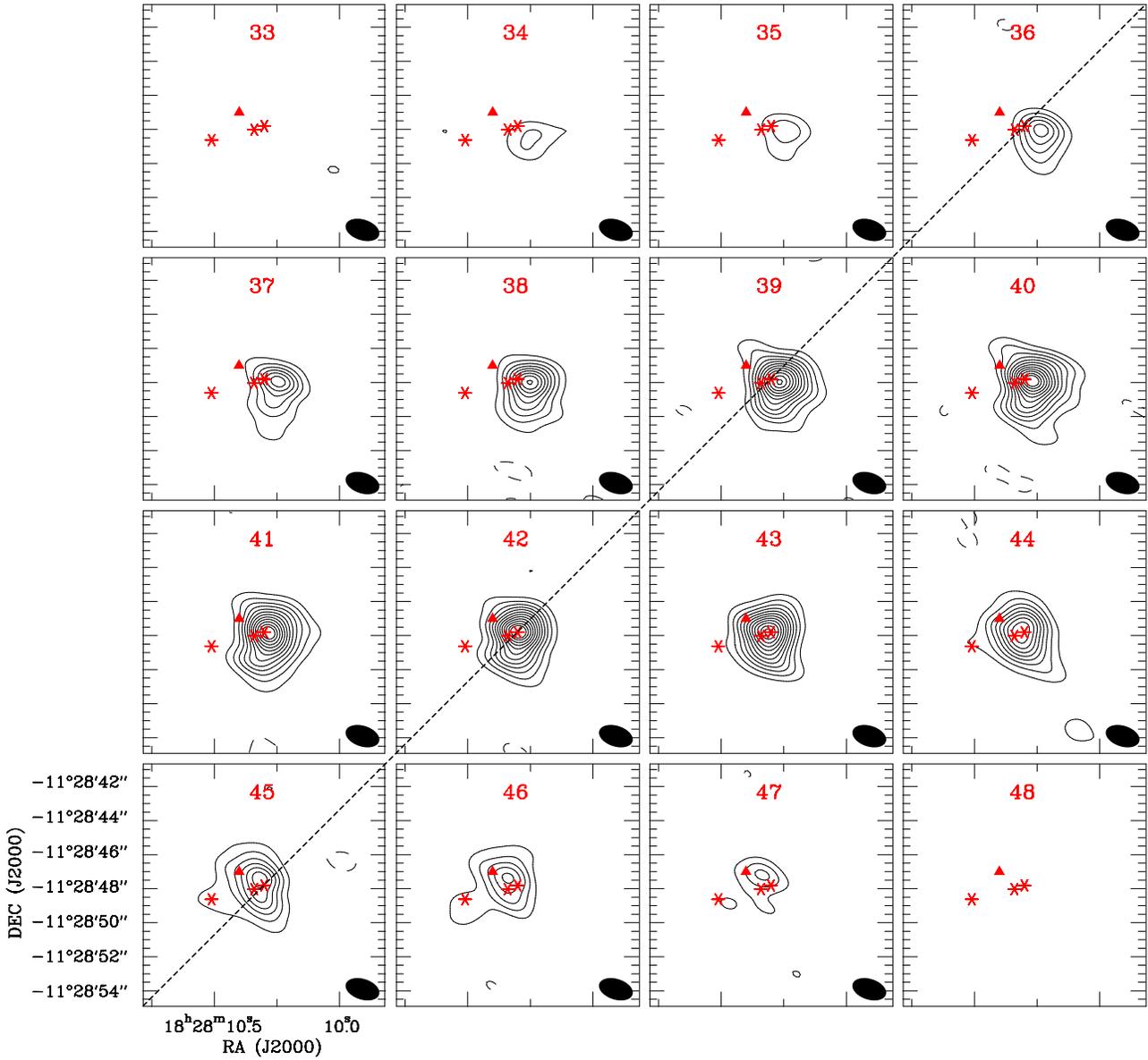}
\vspace{-25mm}\caption{Sample images from $\rm C^{34}S$ and SiO in different
velocity channels. In each panel, the synthesized beam is shown in the
lower-right corner. The red $``\ast"$ and filled triangle indicate
the position of three HII region and $\rm H_{2}O$ maser,
respectively. Central velocities are indicated in each image. Contour levels are all -4, 4, 8, 12, 20, 28, 36, 44
$\sigma$ .... The oblique dashed line is used to determine the peak
positional move of core. (a) $\rm C^{34}S$, 1 $\sigma$ noise level
is 0.1 Jy beam$^{-1}$.} \vspace{3mm}
\end{figure*}

\begin{figure*}
\vspace{0mm}
\includegraphics[angle=270,scale=.95]{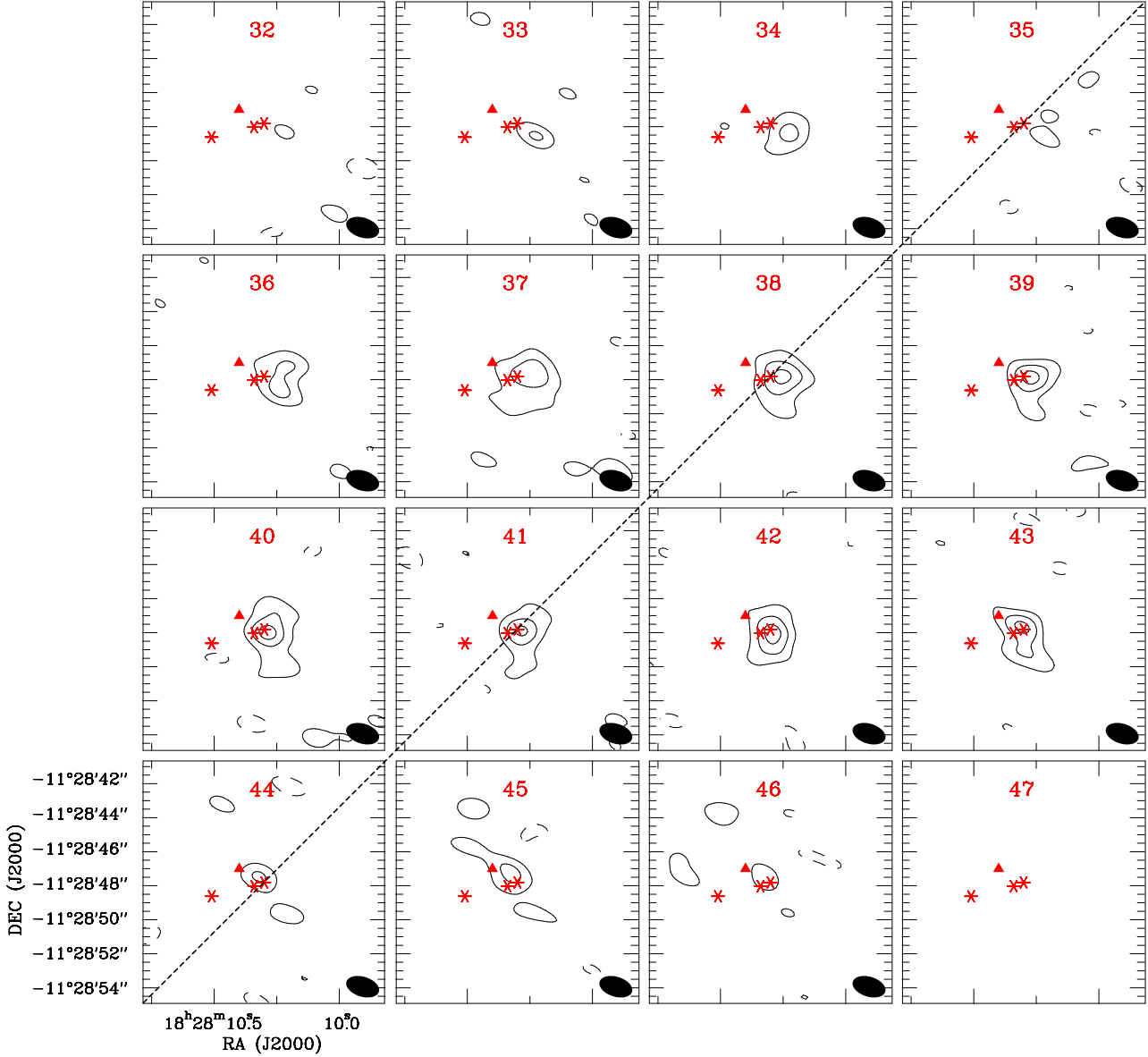}
\vspace{-25mm}\caption{(b) SiO, 1 $\sigma$ noise level is 0.1 Jy
beam$^{-1}$.} \vspace{3mm}
\end{figure*}

\begin{figure*}
\vspace{30mm}
\includegraphics[angle=0,scale=0.8]{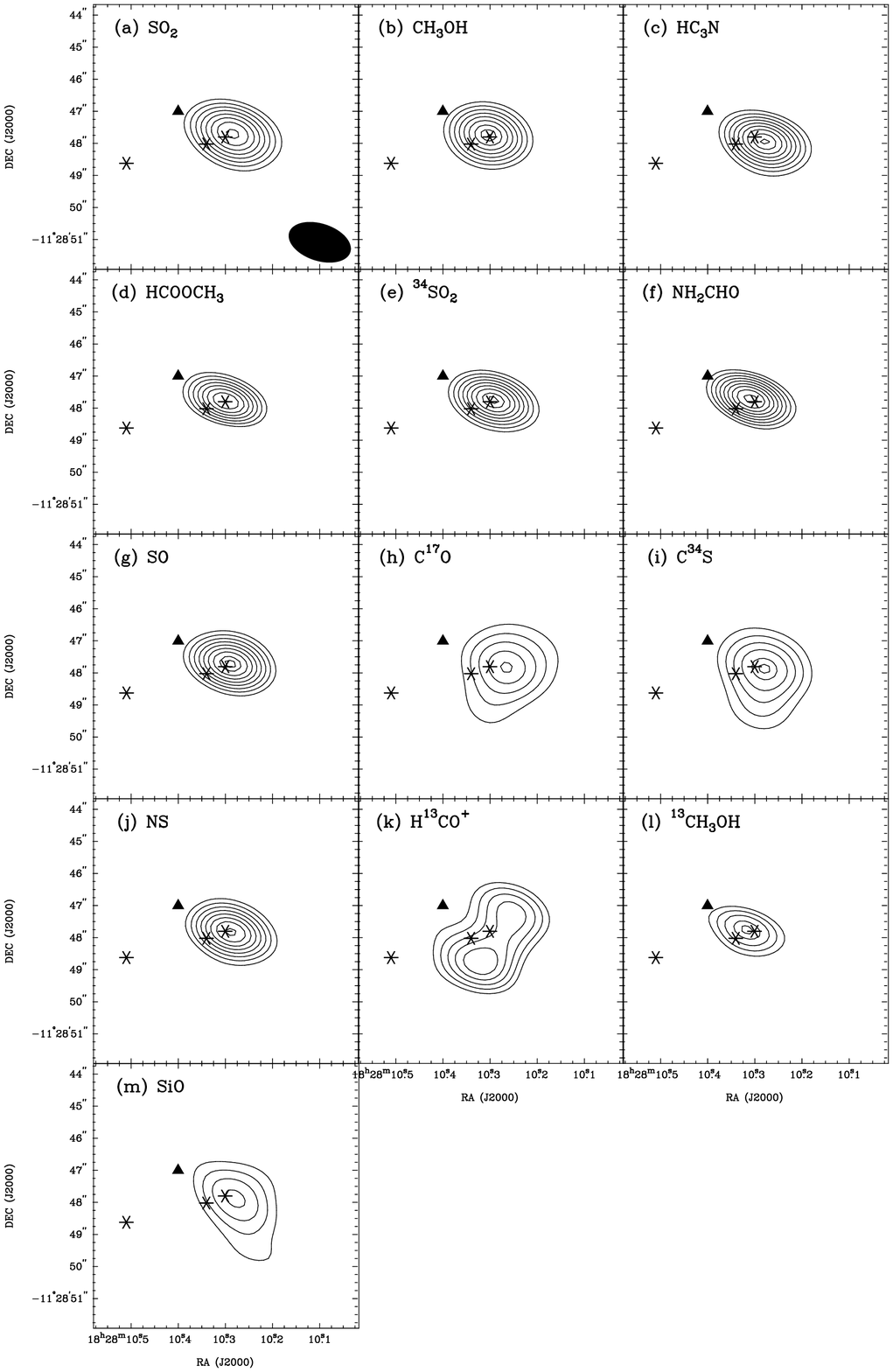}
\vspace{0mm}\caption{The integrated intensity maps of each species.
Contour levels are all 3, 4, 5, 6, 7, 8, 9, 10 $\sigma$ .... The 1
$\sigma$ noise levels (Jy $\rm beam ^{-1}$ km $\rm s^{-1}$) of
various species are presented in Table 1.}
\end{figure*}

To obtain the spatial kinematic distribution of molecules in
G20.08-0.14N, we chose the strongest and unblended lines for the
channel maps if multiple lines for one species were detected. Here
we only show the channel maps of C$^{34}$S and SiO in Figures 3-4.
The contours are made at the velocity channels with interval 1 km
$\rm s^{-1}$, which begin at 4 $\sigma$. In Figures 3-4, we present
the full velocity range of each molecular emission. The three HII
regions are shown with ``$\ast$'' symbols. Since the HII region A is
coincident with the peak position of the 0.9 mm continuum (within
the uncertainty), the symbol of the HII region A can also represent
the peak position of the 0.9 mm continuum. The systemic velocity
($V_{\rm sys}$) for G20.08-0.14N is about 42 km s$^{-1}$
\citep{Plume92,Gal09}. From Figures 3-4, we can see that both
molecules show a velocity gradient across the 0.9 mm continuum along
the northeast-southwest direction. To further determine the peak
move of the 0.9 mm continuum, we made a dashed line across the HII
region A of four panels in Figures 3-4. Also, the peak position of
the continuum above the dashed line is redshifted, while the
blueshifted emission is under the dashed line, further confirmed
that there is a velocity gradient in G20.08-0.14N. According to the
channel maps of each species, we made the integrated intensity map
of each species (Figure 5). In Figure 5, the emission peaks of
CH$_{3}$OH and HCOOCH$_{3}$ are associated with the continuum peak.
The emission peaks of $^{13}$CH$_{3}$OH and $\rm NH_{2}CHO$ are
located to the northeast of the continuum peak, while the emission
peaks of  $\rm HC_{3}N$, $\rm C^{17}O$, $\rm C^{34}S$, $\rm NS$, and
$\rm SiO$ are located to the southwest of continuum peak. Three
sulfur-bearing molecules ($\rm SO_{2}$, $\rm ^{34}SO_{2}$, and $\rm
SO$) are situated to the northwest of continuum peak. Additionally,
the emission of $\rm H^{13}CO^{+}$ shows two molecular cores around
the HII region A.

 The different spatial distribution of molecular gas from
different species is reminiscent of the chemical differentiation
observed in other hot-cores, like Orion-KL. Most of the
nitrogen-bearing molecules peak in the Orion hot core while most of
the oxygen-bearing molecules are found toward the compact ridge
\citep{Blake87,Beuther05,qin10}, similar to what observed in
G19.61-0.23 by \cite{qin10}. At 7 mm, however, a complex
oxygen-bearing molecule (acetone) has been detected toward the Orion
hot core, while two nitrogen-bearing molecules (cyanopolyynes) are
found in the quiescent cold gas of the Orion extended ridge
\citep{goddi09}, similar to what observed in G20.08-0.14N. Further
higher sensitivity and resolution observations especially from the
ALMA are needed to confirm the peak offsets in G20.08-0.14N.

\subsection{Column Densities and Abundances}
In our observations we detected multiple transitions ($>3$) from
$\rm SO_{2}$, $\rm CH_{3}OH$, and $\rm HC_{3}N$, but owing to
spectral blending of lines from $\rm SO_{2}$ and $\rm HC_{3}N$, we
used only $\rm CH_{3}OH$ to calculate the rotation temperature and
the column density by a rotation temperature diagram (RTD). Eight
transitions of $\rm CH_{3}OH$ have been detected in G20.08-0.14N,
containing 5 ground state and 3 vibrationally excited lines. With
the assumptions of the local thermodynamic equilibrium (LTE), lines
being optically thin and gas emission filling the beam, the rotation
temperature and the beam-averaged column density can be determined
by \citep{Goldsmith99,Liu02,qin10}
\begin{equation} \mathit{\rm ln(\it \frac{N_{u}}{g_{u}})}=\rm ln(\it \frac{N_{T}}{Q_{rot}})-\frac{E_{ u}}{T_{ rot}},
\end{equation}
Where $N_{u}$ is the column density of the upper energy level,
$g_{u}$ is the degeneracy factor in the upper energy level, $N_{T}$
is the total beam-averaged column density, $Q_{\rm rot}$ is the
rotational partition function, $E_{u}$ is the upper level energy in
K, and $T_{\rm rot}$ is the rotation temperature. By plotting the
data points from eight transitions of $\rm CH_{3}OH$ according to
Equation (1) and applying least-square fitting for a straight line,
a rotation temperature diagram is shown in Figure 6. The RTD can be
corrected by multiplying $N_{ u}/g_{u}$ by the optical depth
correction factor $C_{\tau}=\tau/(1-e^{\tau})$, where $\tau$ is the
optical depths, which is expressed by (Remijan et al. 2004)
\begin{equation} \mathit{\rm\tau}=\frac{8\pi^{3}S\mu^{2}\nu N_{u}}{3\kappa\Delta\nu T_{\rm rot}g_{u}},
\end{equation}
Where $S$ is the line strength, $\mu$ is the dipole moment,
$\nu$(GHz) is the rest frequency, $\kappa$ is the Boltzmann
constant, and $\Delta\nu$ is the FWHM line width. From the optical
depth-corrected data, we derived a rotational temperature of 105
$\pm$ 29 K and the beam-averaged column density of
(3.1$\pm$2.1)$\times$ $10^{17}$ $\rm cm^{-2}$ for the G20.08-0.14N
region.

\begin{figure}
\vspace{0mm}
\includegraphics[angle=0,scale=.5]{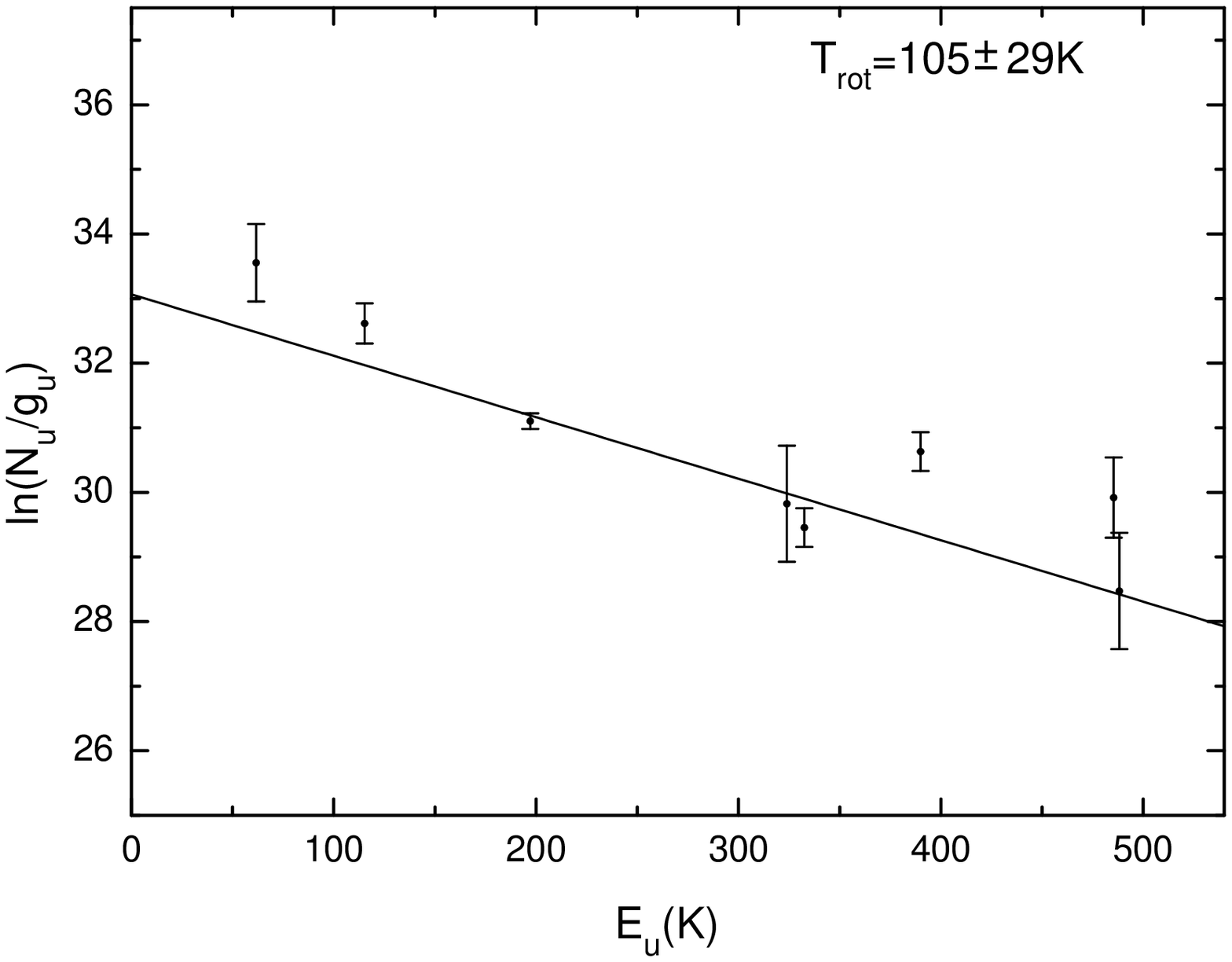}
\vspace{-8mm}\caption{Population temperature diagram of the observed
$\rm CH_{3}OH$ transitions. The vertical bars mark the $\rm
ln$$(N_{\rm_{ u}}/g_{_{\rm u}})$ errors from the integrated
intensities. The linear least-squares fit (solid line) gives a
rotation temperature of 105 $\pm$ 29 K.}
\end{figure}

Following \cite{qin10} (their eq.5), the beam-averaged column
densities of  molecules with less than three transitions detected
could be expressed by
\begin{equation} \mathit{N_{T}}({\rm cm^{-2}})=2.04\times10^{20}\frac{I(T_{\rm rot})}{I(T_{\rm rot})-I_{b}}\times\frac{Q_{\rm rot}e\int Id\upsilon}{\theta_{a}\theta_{b}\nu^{3}S\mu^{2}} ,
\end{equation}
for linear molecules ($T_{\rm rot}$=$E_{u}$).
\begin{equation} \mathit{N_{T}}({\rm cm^{-2}})=2.04\times10^{20}\frac{I(T_{\rm rot})}{I(T_{\rm rot})-I_{b}}\times\frac{Q_{\rm rot}e^{3/2}\int Id\upsilon}{\theta_{a}\theta_{b}\nu^{3}S\mu^{2}} ,
\end{equation}
for symmetric and asymmetric top molecules ($T_{{rot}}$=2/3$E_{u}$).
Where $I(T_{rot}$) and $I_{b}$ are the specific intensities of the
spectrum at $T_{rot}$ and of the background continuum, $\int
Id\upsilon$ is the integrated intensity of the specific transition
in Jy beam$^{-1}$, $\theta_{a}$ and $\theta_{b}$ are the FWHM beam
size in arcsec$^{2}$.

In addition, the fractional abundance of a certain molecule relative
to $\rm H_{2}$ depends on the column densities of the certain
molecule and $\rm H_{2}$ molecule, defined by $f_{\rm H_{2}}=N_{T}
/N_{\rm H_{2}}$ , where $N_{T}$ is the beam-averaged column density.
The optically thin submillimeter dust continuum emission has been
proven to be an effective way to determine $\rm H_{2}$ column
density \citep{Pierce00,Gordon95}. Assuming an average grain radius
of 0.1 micron and grain density of 3 g cm$^{-3}$ and a gas to dust
ratio of 100 \citep{Lis91}, the beam-averaged column density is
given by the formulae \citep{Lis91}
\begin{equation} \mathit{N_{\rm H_{2}}}(\rm cm^{-2})=8.1\times10^{17}\frac{e^{\it h\nu/kT}-1}{Q(\nu)\Omega}(\frac{S_{\nu}}{Jy})(\frac{\nu}{GHz})^{-3},
\end{equation}
where $T$ is the mean dust temperature (K), Q($\nu$) is grain
emissivity at frequency $\nu$, S$_{\nu}$ is the peak intensity of
the continuum, and $\Omega$ is the beam solid angle. Assuming
radiation equilibrium, \cite{Gal09} set the dust temperature to 230
K for G20.08-0.14N. We adopt Q($\nu$) of $4\times10^{-5}$ at 340 GHz
\citep{Lis91} and dust temperature of 230 K for the calculation of
the beam-averaged column density. The intensity of the continuum
peak is 1.33$\pm$0.02 Jy beam$^{-1}$. The derived beam-averaged
column density of $\rm H_{2}$ is  (8$\pm$0.1)$\times$$10^{23}$
cm$^{2}$, which  is reasonably consistent with those
($\sim$$10^{23}$ cm$^{2}$) in NH$_{3}$ \citep{Gal09},
(3$\times$$10^{23}$ cm$^{2}$) in hot core G327.3+0.6 \citep{gibb00}
and (8.4$\times$$10^{23}$ cm$^{2}$) in hot core G19.61+0.23
\citep{qin10}. Therefore, the derived density of $\rm H_{2}$ is
reliable. The fractional abundances of the various species relative
to $\rm H_{2}$ are estimated from the beam-averaged column densities
as shown in the fourth column of Table 2.

\begin{table}
\begin{center}
\tabcolsep 1.3mm\caption{The Parameters Derived From Molecular
Lines}
\begin{tabular}{lcccccccc}
\hline\hline
Molecule   & $T_{\rm rot}$      & $N_{T}$   & $f_{\rm H_{2}}$   \\
         &    (K)       &(cm$^{-2}$)     &     \\
  \hline\noalign{\smallskip}
$\rm SO_{2}$        & 112    & $(2.5\pm0.1)\times10^{16}$   & $(3.2\pm0.1)\times10^{-8}$  \\  
$\rm ^{34}SO_{2}$   & 192    & $(3.2\pm0.2)\times10^{16}$   & $(4.2\pm0.2)\times10^{-8}$  \\  
$\rm SO$            & 143    & $(3.7\pm0.2)\times10^{17}$   & $(4.8\pm0.2)\times10^{-7}$  \\  
$\rm C^{34}S$       & 50     & $(5.1\pm0.1)\times10^{14}$   & $(6.5\pm0.2)\times10^{-10}$    \\  
$\rm NS$            & 71     & $(3.3\pm0.1)\times10^{15}$   & $(4.2\pm0.2)\times10^{-9}$  \\  
$\rm HC_{3}N$       & 307    & $(7.8\pm0.2)\times10^{15}$   & $(1.0\pm0.1)\times10^{-8}$  \\  
$\rm NH_{2}CHO$    & 100    & $(4.4\pm0.5)\times10^{14}$   & $(5.7\pm0.3)\times10^{-10}$ \\  
$\rm C^{17}O$       & 33     & $(7.0\pm0.2)\times10^{17}$   & $(9.0\pm0.2)\times10^{-7}$   \\  
$\rm SiO$           & 75     & $(1.8\pm0.1)\times10^{15}$   & $(2.3\pm0.2)\times10^{-9}$   \\  
$\rm H^{13}CO^{+}$  & 28     & $(2.2\pm0.2)\times10^{14}$   & $(2.9\pm0.2)\times10^{-10}$   \\  
$\rm CH_{3}OH$      & 105$\pm$29  & $(3.1\pm2.1)\times10^{17}$   & $(4.0\pm2.7)\times10^{-7}$  \\  
$\rm ^{13}CH_{3}OH$ & 170    & $(3.5\pm0.7)\times10^{15}$   & $(4.5\pm0.8)\times10^{-9}$  \\  
$\rm HCOOCH_{3}$    & 165    & $(3.5\pm0.3)\times10^{17}$   & $(4.5\pm0.4)\times10^{-7}$  \\  

\hline
\end{tabular}\end{center}
\vspace{0mm}
\end{table}

\section{DISCUSSION}
\subsection{Kinematics}
Recently, through numerical simulation, \cite{Vazquez-Semadeni09}
suggest that the formation of massive stars is coincident with
large-scale collapse, while low-mass and intermediate-mass stars are
associated with isolated accretion flows. Previous observations with
the SMA in hot core molecules (CH$_{3}$CN, OCS, and SO$_{2}$) and
the VLA in NH$_{3}$ show that G20.08-0.14N is surrounded by a
smaller scale and a large scale accretion flows \citep{Gal09}. The
smaller scale accretion flow is across the HII region A, which may
be resupplied by the large scale accretion flow. The channel maps of
SiO and C$^{34}$S in Figures 3-4 exhibit a velocity gradient along
the SW-NE, suggesting that this is  an inflow or rotation motions
around the HII region A of G20.08-0.14N.

To explore whether these  are two accretion flows in G20.08-0.14N,
we made the position-velocity (P-V) diagrams (Figure 7) of SiO and
C$^{34}$S lines across the HII region A with cuts at
P.A.=45$^{\circ}$ and 135$^{\circ}$. Since SiO is more easily
affected by the excitation conditions, the PV diagrams of SiO
present the complex pattern, which contain several velocity
components.  To analyze in detail each component, we divide the
whole component into six region. Each has been designated
alphabetically: Regions D-I.  From Figure 7 (a) and (b), we find
two velocity components in Region D and Region I, which are not seen
in the perpendicular direction P.A.=135$^{\circ}$. The velocity
component of Region D is from 32.5 km s$^{-1}$ to 36.0 km s$^{-1}$,
while the velocity component of Region I is from 47.0 km s$^{-1}$ to
51.7 km s$^{-1}$. By comparing with previous observations in
CH$_{3}$CN, OCS, SO$_{2}$ and NH$_{3}$ \citep{Gal09}, we find that
the two velocity components of SiO detected in Region D and Region I
do not belong to the velocity ranges of the accretion flows of
\cite{Gal09}.  Since SiO is considered as a good tracer of
outflow (e.g., Schilke et al. 1997b; Gueth \& Guilloteau 1999;
Cesaroni et al. 1999; Beuther et al. 2005), we suggest that the two
velocity components of SiO may arise from an outflow  in the
NE-SW direction.  To confirm the interpretation of the two
velocity components in terms of outflow, we made the integrated
intensity maps for each component as shown in Figure 1. In Figure 1,
the blueshifted and redshifted components are presented as blue and
red contours. The map clearly shows NE-SW bipolar components
centered at the peak position of the continuum. Hence, we suggest
that the velocity gradients seen in Region D and Region I are caused
by the bipolar outflow motions. The blueshifted and redshifted lobes
of the outflow are in a straight line at the NE-SW direction.
Klaassen \& Wilson (2008) suggested an outflow with SiO line
profiles in G20.08-0.14N, while \cite{Gal09} did not detect the
outflow in CO lines. Here, we give for the first time the direction
of the outflow. A dynamic time scale of the outflow can be
determined by $t_{d}=r/v$, where $v$ is the maximum flow velocity
relative to $V_{\rm sys}$, and $r$ is the length of the begin-to-end
flow extension for each lobe.  From Figure 1, we derived that
the lengths of the blueshifted and redshifted lobe S are 0.15 pc and
0.36 pc, respectively. The obtained average dynamical timescale is
about 2.6 $\times$ $10^{4}$ yr. An H$_{2}$O is located at the
redshifted lobe of the outflow, which may be excited by the shocks
from the outflow. The HII region A is situated at the center of the
outflow and may drive the collimated outflow.

\begin{figure*}
\vspace{0mm}
\includegraphics[angle=180,scale=.85]{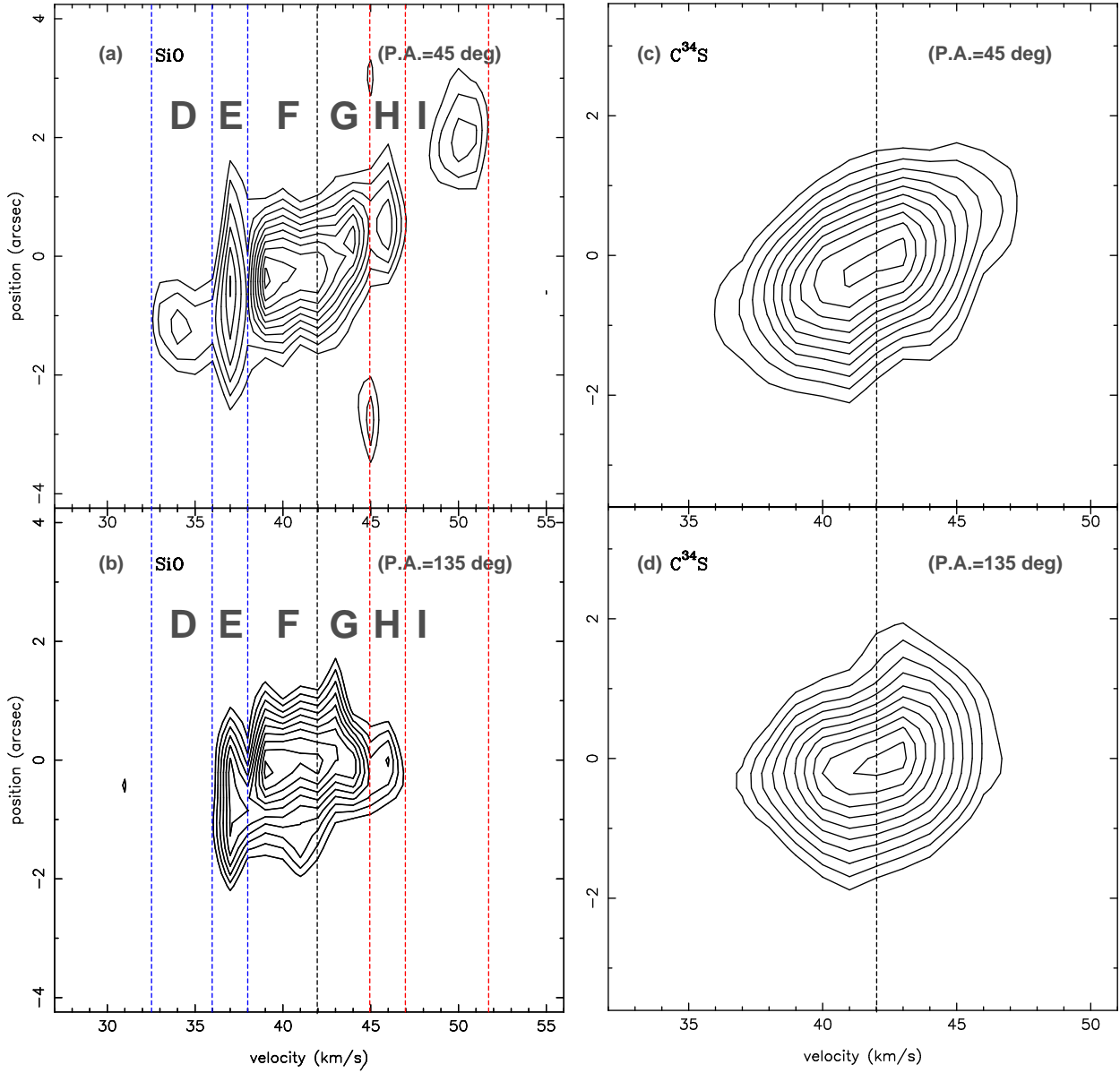}
\vspace{-30mm} \caption{PV diagrams of SiO and C$^{34}$S. The upper
diagrams are for a cut at P.A.=45 (NE-SW), while the below diagrams
are at P.A.=135 (NW-SE). Cuts are across the position of the HII
region A. The black dashed line indicates $V_{\rm sys}$. Other blue
and red lines are used to mark the beginning and end of each
velocity component.}
\end{figure*}

Moreover, the velocities of Region F and Region G at peaked position
are 39.0 km s$^{-1}$ and 44 km s$^{-1}$, respectively, which both
may have a velocity gradient of  $\sim$ 2.0 km s$^{-1}$ with respect
to $V_{\rm sys}$. The velocity gradient is well coincident with the
inward velocity found in the NH$_{3}$ absorption line at smaller
scale \citep{Gal09}. While the velocity of both Region E and Region
H with respect to $V_{\rm sys}$ is $\sim$ 4.0 km s$^{-1}$, which is
well associated with the rotation velocity at larger scale
\cite{Gal09}. Additionally, from Figure 7 (c) and (d), we find also
that the PV diagram of C$^{34}$S show a velocity gradient at
P.A.=135$^{\circ}$  and P.A.=45$^{\circ}$. C$^{34}$S is used to
trace the Keplerian rotation \citep{Beuther09}. \citep{Gal09} also observed a velocity gradient in molecular lines at P.A.=45$^{\circ}$,
which they interpret as rotation in a torus/disk whose plane is oriented NE-SW. Our measurements however
provide clear indication of an outflow along the same NE-SW direction.
Therefore, we favor a model where the detected velocity gradients at P.A.=135$^{\circ}$
degrees are caused by rotation motions (and possibly infall),in the perpendicular direction of the outflow.
However, in this model we can not explain the velocity gradients detected at P.A.=45$^{\circ}$ degrees in the Regions E-H of Figure 7 (a) and Figure 7 (c).

\subsection{Chemistry}
From Table 1, we can see that 11 species were detected. The emission
peaks for different molecules are located at different positions in
the core (Figure 5), hence each molecule may be produced via a
different mechanism. The individual molecules are discussed in the
following.

\subsubsection{Sulfur-bearing Molecules}
The sulphur chemistry is of specific interest because of its rapid
evolution in warm gas and since the abundances of sulphur-bearing
species increase significantly with temperature, both by ice
evaporation and by shock interaction.

\it{Sulphur dioxide} \rm(SO$_{2}$)--Eight transitions of SO$_{2}$
were detected, containing 4 ground-state and 4 vibrationally excited
lines. SO$_{2}$($19_{_{1,19}}$-$18_{_{0,18}}$) at 346.65216 GHz is
strongest and unblended line. Because SO$_{2}$ is an inorganic
asymmetric molecule, adopting a rotation temperature of $T_{\rm
rot}$=2/3$E_{u}$, we estimated that the column density and abundance
of SO$_{2}$ are $(2.5\pm0.1)\times10^{16}$ and
$(3.2\pm0.1)\times10^{-8}$, respectively. The derived abundance of
SO$_{2}$ is consistent with those in Sgr B2, G29.26 and G19.61-0.23
\citep{Nummelinl00,Beuther09,qin10}, but larger than that in Orion
KL \citep{Beuther09}.  It has been recently suggested that the
Orion-KL hot core may be only a pre-existing density enhancement
heated from the outside by shocks \citep{Zapata11, Goddi11}. Via
the shock-induced chemical models, \cite{Hartquist80} estimated that
abundance of SO$_{2}$ is $3.7\times10^{-11}$.  The highest
fractional abundance relative to H$_{2}$ in G20.08-0.14N can be
explained by grain surface chemistry, not by shock interaction of a
molecular outflow with the ambient dense gas. Two transitions of the
$^{34}$SO$_{2}$ isotopologue were detected.  In Orion, SO$_{2}$
dominates the appearance of the millimeter-wave spectrum
\citep{sch97a}, accounting for approximately 28\% of all the
detected lines. Because of its asymmetric geometry, it has a rich
spectrum of lines which are typically very strong because of the
large abundance and high dipole moment of the molecule. In
G20.08-0.14N, the 10 transition lines of the detected 41 transition
lines  belong to SO$_{2}$ and its isotopologue, which is similar to
that in Orion.

\it{Sulphur monoxide} \rm(SO)--Sulphur monoxide was detected in
($10_{_{11}}$-$10_{_{10}}$) transition at 336.55375 GHz. The derived
column density and abundance is $(3.7\pm0.2)\times10^{17}$ and
$(4.8\pm0.2)\times10^{-7}$, respectively. \cite{Charnley97} has
investigated the formation  of SO and SO$_{2}$. When t$\leq$10$^{4}$
yr, the evolution of the sulfur chemistry in the models shows that
sulfur monoxide is produced by reaction SH+O$\rightarrow$SO+H; When
t$\geq$2$\times$10$^{4}$ yr, the reaction S+OH$\rightarrow$SO+H is
also important at 100 K, along with S+O$_{2}$$\rightarrow$SO+O. For
increasing core temperature, reaction S+O$_{2}$$\rightarrow$SO+O
dominates. Considering the SO$_{2}$ abundance of
$(3.2\pm0.1)\times10^{-8}$, we obtained the SO/SO$_{2}$ abundant
ratio of 15$\pm$ 0.2 in G20.08-0.14N, which can be explained by the
evaporated mantles model without O$_{2}$ injection at 200 K. From
the model we estimated that the age of the G20.08-0.14N hot core is
about 8$\times$10$^{3}$ yr, which is less than the dynamical
timescale of the outflow identified by SiO. In Figure 5, the
emission peaks of SO is located to the northwest of the continuum
peak, which is perpendicular to the direction of the outflow. Hence,
although the outflow can excite the formation of SO from the
timescales, SO may be produced by reaction SH+O$\rightarrow$SO+H.
After some time, we may detect the SO excited by the outflow.

\it{Carbon monosulfide} \rm(CS). The rare carbon-sulfur isotopologue
$\rm C^{34}S$ ($7$-$6$)  is detected at 337.39646 GHz. The estimated
column density and fractional abundance of $\rm C^{34}S$ is
$(5.1\pm0.1)\times10^{14}$ and $(6.5\pm0.2)\times10^{-10}$,
respectively.  The derived column density of $\rm C^{34}S$ is
similar to the one observed in Orion Kl \citep{Beuther09}.
\cite{Beuther09} considered that the formation of $\rm C^{34}S$ can
be successfully explained by gas chemistry models. Moreover, the PV
diagrams of $\rm C^{34}S$ show the inflow and rotation motions,
further confirming that $\rm C^{34}S$ should be a better trace of
rotational disk at very early evolutionary stages.

\subsubsection{Nitrogen-bearing Molecules}
Theoretical models and observations suggest that the
nitrogen-bearing molecules have higher gas temperatures and lower
fractional abundances on a timescale of$\sim$10$^{5}$ yr
\citep{Blake87,Rodgers01,Miao95,Kuan96,qin10}. In the G20.08-0.14N
hot core, \cite{Gal09} detected NH$_{3}$ and CH$_{3}$CN molecules.
They derived that the fractional abundances of both molecules are
about 5$\times$10$^{-7}$ and 5$\times$10$^{-9}$ to
2$\times$10$^{-8}$.  NH$_{3}$ and CH$_{3}$CN are the best
tracers of hot cores \citep{Wilner94,Wright96,Wilson00}. NH$_{3}$
may origin from the evaporation of grain mantles \citep{Pauls83},
which drives {\bf a} nitrogen-rich chemistry and produces much
complex nitrogen-bearing molecules \citep{Caselli93}. Here, we
detected three nitrogen-bearing molecules, including cyanoacetylene,
nitrogen sulfide, and formamide toward the G20.08-0.14N hot core.

\it{Cyanoacetylene} \rm(HC$_{3}$N). We detected the two ground-state
and three vibrationally excited state lines of HC$_{3}$N. HC$_{3}$N
($38$-$37e$) at 346.45573 GHz is strongest and unblend. Using this
transition line, we obtained the HC$_{3}$N column density and
fractional abundance of $(7.8\pm0.2)\times10^{15}$ and
$(1.0\pm0.1)\times10^{-8}$, respectively. The derived fractional
abundance of HC$_{3}$N is one order of magnitude larger than that in
Orion KL and Sgr B2 \citep{Beuther09,Nummelinl00}. The fractional
abundance of HC$_{3}$N with respect to the NH$_{3}$ is 0.02.

\it{Nitrogen sulfide} \rm(NS) Nitrogen sulfide at 346.22116 GHz is
the first detected in the $J=15/2-13/2$ transition. NS is a
relatively simple species, which provides a good test for coupled
chemistry models of nitrogen and sulfur. The derived column density
of $\rm NS$ is $(3.3\pm0.1)\times10^{15}$. The fractional abundance
of NS relative to molecular hydrogen is $(4.2\pm0.2)\times10^{-9}$,
which is larger than about 10$^{-11}$ found by some recent gas-phase
chemistry models developed for quiescent clouds \citep{Lee96}. In
addition, \cite{Viti01} predicted that the abundance of NS is
significantly enhanced by shock. NS does appear to follow the
distribution of SO and CH$_{3}$OH in TMC-1 \citep{McGonaglel97}. In
G20.08-0.14N, the emission peak of NS is similar to the distribution
of HC$_{3}$N and SiO. The emission of the SiO shows the outflow.
Hence, we suggest that the formation of NS may be related to the
outflow of G20.08-0.14N.

\it{Formamide} \rm(NH$_{2}$CHO) We also detected NH$_{2}$CHO at
336.13688 GHz. Assuming a rotation temperature of $T_{rot}$ =
2/3$E_{u}$, the estimated column density of NH$_{2}$CHO is
$(4.4\pm0.5)\times10^{14}$, which is close to values in Orion KL
\citep{sch97a} and in G19.61-0.23 \citep{qin10}. Because the
abundance of NH$_{2}$CHO is 10 times less than that of the HNCO in
G19.61-0.23, \cite{qin10} suggest that NH$_{2}$CHO can be only
formed by successive hydrogenation of the HNCO on grain surface.
\cite{Tielens97} considered that NH$_{2}$CHO is most likely formed
by atom addition to HCO on grain mantles and then evaporated.
Moreover, \cite{Bernstein95} found  that NH$_{2}$CHO is formed
through UV photolysis and heating of a
H$_{2}$O:CH$_{3}$OH:CO:NH$_{3}$=100:50:10:10 mixture. If NH$_{2}$CHO
is formed via UV emission, \cite{gibb00} suggested that the
abundance of CH$_{3}$OH is higher enough than that of NH$_{3}$. For
G20.08-0.14N, the peak  position of NH$_{2}$CHO with respect to
the continuum is similar to the H$^{13}$CO$^{+}$, and the abundance
ratio of CH$_{3}$OH is lower than that of NH$_{3}$, indicating that
the formation of NH$_{2}$CHO may be similar to the suggestion
of \cite{Tielens97}.

\subsubsection{Oxygen-bearing Molecules}
Comparing with the nitrogen-bearing molecules, the oxygen-bearing
molecules have lower gas temperatures and higher fractional
abundances on a timescale of$\sim$10$^{4}$ yr (Blake et al. 1987;
Rodgers \& Charnley 2001,2003; Qin et al. 2010). CH$_{3}$OH and
HCOOCH$_{3}$ were detected in G20.08-0.14N, which are the typical
oxygen-bearing molecules. CH$_{3}$OH is considered as the precursor
of the oxygen-bearing molecules, which  drives an oxygen-rich
chemistry and produce new molecules such as HCOOCH$_{3}$.

\it{Methanol} \rm(CH$_{3}$OH). Methanol has been detected in
CH$_{3}$OH and its isotopologue $^{13}$CH$_{3}$OH. There are eight
transitions of the CH$_{3}$OH with upper level energies of 61-489 K
including 5 ground state and 3 vibrationally excited lines. Through
the RTD fit, the derived rotation temperature of the CH$_{3}$OH is
larger than 100 K, suggesting that they originated from warm gas
environments. In addition, the column density and fractional
abundance of CH$_{3}$OH are $(3.1\pm2.1)\times10^{17}$ and
$(4.0\pm2.7)\times10^{-7}$, respectively. The obtained fractional
abundance of CH$_{3}$OH is one order of magnitude lower than that in
Orion KL and Sgr B2 \citep{Beuther09,Nummelinl00}, but is consistent
with that in G19.61-0.23 \citep{qin10} and IRAS 20126+4104
\citep{xu12}.  \cite{qin10} and \cite{xu12} suggest that the higher
fractional abundance of CH$_{3}$OH with respect to H$_{2}$ can be
explained by grain surface chemistry. The grain surface chemistry
models predict that the abundance of CH$_{3}$OH is larger than
10$^{-8}$ \citep{van der00}, while the gas chemical models predict
an abundance of$<10^{-9}$ \citep{Lee96}. In G20.08-0.14N, the higher
fractional abundance of CH$_{3}$OH cannot be interpreted simply by
gas-phase chemical reactions, and may be originated from grain
surface chemistry. The $^{12}$C/$^{13}$C ratio from the CH$_{3}$OH
and $^{13}$CH$_{3}$OH is $\sim$ 89. \cite{Milam05} concluded that
the $^{12}$C/$^{13}$C ratio obtained from CO  indicates a
gradient with Galactic distance of $^{12}$C/$^{13}$C=5.41D+19.03,
where D is distance from the Galactic center in kpc. Using the
$^{12}$C/$^{13}$C ratio from the CH$_{3}$OH, we derived that the
distance of G20.08-0.14N is 12.9 kpc, which is close to values
reported by \cite{Fish03} and \cite{Anderson09}.

\it{Silicon monoxide} \rm(SiO). SiO J=8-7(v=0) was detected at rest
frequencies of 347.33063 GHz with  upper level energy of 75 K. The
column density and fractional abundance derived from a single line
of SiO are $(1.8\pm0.1)\times10^{15}$ and
$(2.3\pm0.2)\times10^{-9}$, respectively. The derived column density
of SiO is an order of magnitude higher than that of Orion KL and
G29.96 \citep{Beuther05}. One explanation for the SiO abundance
enhancements in shocked gas is that SiO is produced by destruction
of grain cores in shocks \citep{Caselli97,sch97b}. Another
possibility is that SiO is embedded in icy grain mantles, which are
evaporated. Hence, we suggest that SiO may be produced by
destruction of grain cores in the shocks of the outflow.

\it{Formyl ion} \rm(HCO$^{+}$) and \it{Methyl formate}
\rm(HCOOCH$_{3}$). We detected the ($4$-$3$) transitions of the
H$^{13}$CO$^{+}$ isotopologue. The estimated column density and
fractional abundance  of H$^{13}$CO$^{+}$ are
$(2.2\pm0.2)\times10^{14}$ and $(2.9\pm0.2)\times10^{-10}$. In
Figure 5, the emission of HCO$^{+}$ shows {\bf an} elongated
structure with two cores.  Since the HCO$^{+}$ is blended with
the line from CH$_{3}$CH$_{2}$CN at 346983.8 MHz, the latter could
be responsible for the complex structure of HCO$^{+}$. HCOOCH$_{3}$
is a heavy asymmetric rotor with hindered internal rotation of the
methyl group. Two transitions of HCOOCH$_{3}$ were detected with the
relatively weak emission. Using the transition of HCOOCH$_{3}$ at
347.47824 GHz, we obtained the column density and fractional
abundance  of $(3.5\pm0.3)\times10^{17}$ and
$(4.5\pm0.4)\times10^{-7}$, respectively. Our obtained  column
density of HCOOCH$_{3}$ is consistent with that of Sgr B2(N)
\citep{Liu01}, but larger than that {\bf($\sim$10$^{-8}$)} produced
by gas phase chemistry.

In the Orion KL hot core,  because the hot molecular gas is not
associated with any self-luminous millimeter, radio, or embedded
infrared source, \cite{Goddi11} and \cite{Zapata11} suggested that
the Orion-KL hot core may be only a pre-existing density enhancement
heated from the outside by shocks from a molecular outflow. On the
other hand, in G20.08-0.14N the molecular gas is associated with a
millimeter source and an HII region and clearly shows star formation
activity. This indicates that the G20.08-0.14N hot core is heated by
a protostar forming at its center. Based on the chemical models
discussed here, we concluded that the age of the G20.08-0.14N hot
core should be about 10$^{4}$-10$^{5}$ years.

\section{SUMMARY}
The Submillimeter Array observations toward the high-mass
star-forming region G20.08-0.14N are presented in the submillimeter
continuum and in the molecular line transitions. From the SMA data,
0.9 mm continuum emission reveals an extended structure which is
associated with an $\rm H_{2}O$ maser and three UC HII and HC HII
regions. Forty-one molecular transitions were detected related to 11
molecular species, including $\rm SO_{2}$, $\rm SO$, $\rm C^{34}S$,
$\rm NS$, $\rm C^{17}O$, $\rm SiO$, $\rm CH_{3}OH$, $\rm HC_{3}N$,
$\rm H^{13}CO^{+}$, $\rm HCOOCH_{3}$, $\rm NH_{2}CHO$ and their
isotopic species. In addition,  six molecular transitions are
unidentified.  10 transition lines of the detected 41 transition
lines belong to SO$_{2}$, which dominates the appearance of the
submillimeter-wave spectrum. The channel maps of $\rm C^{34}S$ and
$\rm SiO$  show velocity gradients. In their PV diagram,
$\rm C^{34}S$ emission show  rotation motions, while $\rm SiO$
not only present two rotation motions in smaller and larger scales
respectively, but also reveal for the first time  a collimated
outflow along the NE-SW direction. The average dynamical timescale
of the outflow is about 2.6 $\times$ $10^{4}$ yr. An HII region is
situated in the central position of the outflow, which may drive the
collimated outflow. Eight transitions of $\rm CH_{3}OH$ are
unblended, we derived the rotational temperature and the column
density of 105 K and 3.1 $\times$ $10^{17}$ $\rm cm^{-2}$ for  $\rm
CH_{3}OH$ lines, respectively, further indicating that a hot core
coincides with G20.08-0.14N.  The hot core is heated by a
protostar at its center with the age of about 10$^{4}$-10$^{5}$. The
emission peaks of different molecules are located at different
positions of the hot core. By comparing the abundances of different
species with chemical models and previous observations of other hot
cores, we concluded that each species may be produced with a
different mechanism. Nitrogen sulfide (NS) is for the first time
detected in G20.08-0.14N.

\section*{Acknowledgments} We thank an anonymous referee for very useful suggestions; We thank also the SMA staff for the
observations. Jin-Long Xu's research is in part supported by 2011
Ministry of Education doctoral academic prize. Also supported by the
young researcher grant of national astronomical observatories,
Chinese academy of sciences.

\label{lastpage}
\end{document}